\documentclass[twocolumn,english,aps,prl]{revtex4}
\usepackage[T1]{fontenc}
\usepackage{amsmath}
\usepackage{graphicx}
\usepackage{amssymb}
\usepackage{esint,color}

\usepackage{babel}
\usepackage{dsfont}

\renewcommand{\Im}{{\rm Im}}
\newcommand{\Tr}{{\rm Tr}}
\newcommand{\rd}{{\rm d}}
\newcommand{\kb}{k_{\rm B}}

\begin{document}

\title{Heat flux splitter for near-field thermal radiation}

\author{P. Ben-Abdallah}
\email{pba@institutoptique.fr} 
\affiliation{Laboratoire Charles Fabry, Institut d'Optique, CNRS, Universit\'{e} Paris-Sud, Campus
Polytechnique, RD128, 91127 Palaiseau Cedex, France.}

\author{A. Belarouci}
\affiliation{Laboratoire Nanotechnologies Nanosyst\`{e}mes (LN2)-CNRS UMI-3463, Universit\'{e} de Sherbrooke, 3000 Boulevard de l{'}Universit\'{e}, Sherbrooke, J1K 0A5, Qu\'{e}bec, Canada.}

\author{L. Frechette}
\affiliation{Laboratoire Nanotechnologies Nanosyst\`{e}mes (LN2)-CNRS UMI-3463, Universit\'{e} de Sherbrooke, 3000 Boulevard de l{'}Universit\'{e}, Sherbrooke, J1K 0A5, Qu\'{e}bec, Canada.}

\author{S.-A. Biehs}
\affiliation{Institut f\"{u}r Physik, Carl von Ossietzky Universit\"{a}t,
D-26111 Oldenburg, Germany.}

\date{\today}

\begin{abstract}
We demonstrate the possibility to efficiently split the near-field heat flux exchanged between 
graphene nano-disks by tuning their doping. This result paves the way for the developement of an active 
control of propagation directions for heat fluxes exchanged in near-field throughout integrated 
nanostructures networks.
\end{abstract}

\maketitle

The control of electric currents in solids is at the origin of modern computer technology which has 
revolutionized our daily life. Until the 2000s no thermal counterpart had been developed to control the 
flow of heat at the nanoscale in a similar manner. In 2006, a step forward in this direction has been 
done by Li et al.~\cite{Casati1} when they introduced the first concept of a thermal transistor for controlling 
heat fluxes carried by phonons through solid segments. Later, several prototypes of phononic thermal 
logic gates~\cite{BaowenLi2} as well as thermal memories~\cite{BaowenLi3} were developed in order to process 
information by means of the heat fluxes carried by phonons~\cite{Sklan2015}. Besides, different solid-state 
thermal diodes were conceived~\cite{BaowenLi2004,Chang,Segal,Cao,BaowenLiEtAl2012} allowing for 
rectifying these fluxes in asymmetric solid segments.

Very recently, there has been a fast growing interest in developing active functionalties to 
manage heat transfers by radiation rather than by conduction between contactless solids. Since 
2010 several radiative thermal rectifiers~\cite{OteyEtAl2010,Iizuka,Fan,BasuFrancoeur2011,PBA_APL,NefzaouiEtAl2013,Zhang2,Huang,Zhu2,Dames} 
have been proposed using spectrally selective nanostructures and phase-changes materials.
In 2014 and 2015 a radiative analog of a transistor was suggested theoretically which allows 
for switching, modulating and amplifying the heat flux exchanged both in the near-field~\cite{PBA_PRL2014} or far-field
regime~\cite{PBA_Karl}. Furthermore, a concept of a radiative thermal memory was introduced 
working in the far-field~\cite{Slava} and the near-field regime~\cite{DyakovMemory}. 
Such devices open the way for new perspectives concerning the development of contactless thermal circuits 
intended for an active thermal management with photons instead of electrons or phonons.
Finally, the thermal diode concept based on phase-change materials has already been tested, experimentally, 
in the far-field regime~\cite{ItoEtAl2014} in 2014. A review of these recent developments can be found
in Ref.~\cite{inv_rev}.

In this Letter, we introduce the concept of a heat flux splitter which allows us to tune the direction of propagation of 
heat flux exchanged in the near-field. To demonstrate the operating modes of this heat splitter we consider 
a set of three graphene nano-disks in mutual interaction. We show that the heat flux exchanged between 
these nano-disks cannot only be splitted equally in two predefined directions as for a 50:50 beam splitter, which is
trivial, but it can also be oriented in mostly one predefined direction acting like a 99:1 beam splitter
by an appropriate tuning of the Fermi levels in the graphene nano-disks. Since this tuning can be achieved by electrical gating, for 
instance, we thus demonstrate, in particular, that the direction of propagation of the radiative heat flow can be 
dynamically controlled in nano-architectures by electrical means.

\begin{figure}[Hhbt]
\includegraphics[scale=0.3]{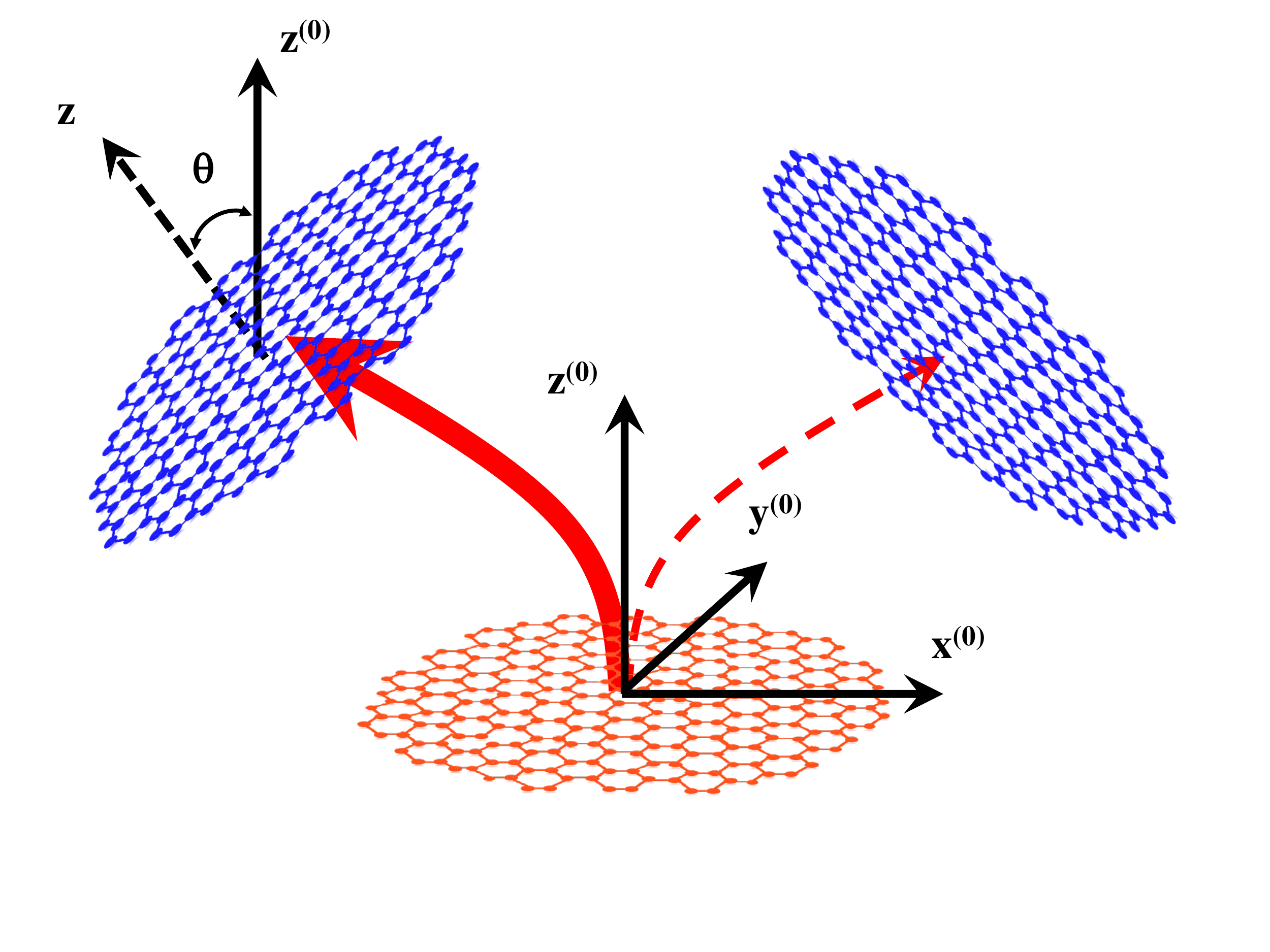}
\caption{Sketch of the heat flux splitter. Three graphene disks with different Fermi levels exchange thermal energy 
in the near-field through many-body interactions. The magnitude of heat flow can be controlled by an appropriate 
tuning of graphene Fermi level.} 
\end{figure}

The geometrical setup of the heat flux splitter is sketched in Fig.~1. Three nano-disks of graphene are distributed 
inside a transparent dielectric of permittivity $\epsilon_h$. Each nano-disk is located at its mass center position $r_{i}$ and its orientation
with respect to the canonical frame $(x^{(0)},y^{(0)},z^{(0)})$ which is determined by the rotation matrix $\mathbf{R}_i$. 
Those nano-disks are supposed to be held at different 
temperatures $T_{i}$. Furthermore, for the sake of clarity, we assume here that the nano-disks are small enough 
compared with the smallest thermal wavelength $\lambda_{T_{i}} = c\hbar/(\kb T_{i})$ ($c$ is the vacuum light
velocity, $2 \pi \hbar$ is Planck's constant, and $\kb$ is Boltzmann's constant) so that they can be modeled 
by simple radiating electrical dipoles. The polarizability of graphene nano-disks in the $x^{(0)}$ and $y^{(0)}$ directions 
is described by a Lorentzian line shape~\cite{de Abajo}
\begin{equation}
  \alpha_{i}(\omega)=3\frac{c^3\kappa_{r,i}}{2\omega_{p,i}^2}\frac{1}{\omega_{p,i}^2-\omega^2-i\kappa_i\omega^3/\omega_{p,i}^2}\label{Eq:Polarizability},
\end{equation}
where $\omega_{p,i}$ is the plasma frequency of the $i^{th}$ disk, $\kappa_i$ is its decay rate, and $\kappa_{r,i}$ 
is the radiative contribution to $\kappa_i$. Along the $z^{(0)}$ direction this polarizabiliy vanishes so that, in 
the canonical basis, the polarizability tensor reads $\boldsymbol{\alpha}_i^{(0)} = {\rm diag}(\alpha_i,\alpha_i,0)$. 
Thus, the polarizability tensor for the oriented nano-disk is simply described in the canonical frame by the 
matricial product
\begin{equation}
  \boldsymbol{\alpha}_{i}(\omega)=\mathbf{R}_i\boldsymbol{\alpha_i}^{(0)}(\omega)\mathbf{R}_i^{-1}\label{Eq:Polarizability2},
\end{equation}
where $\mathbf{R}_i$ denotes the rotation matrix related to the $i^{th}$ disk (see Fig.~1).

\begin{figure}[Hhbt]
  \includegraphics[scale=0.35]{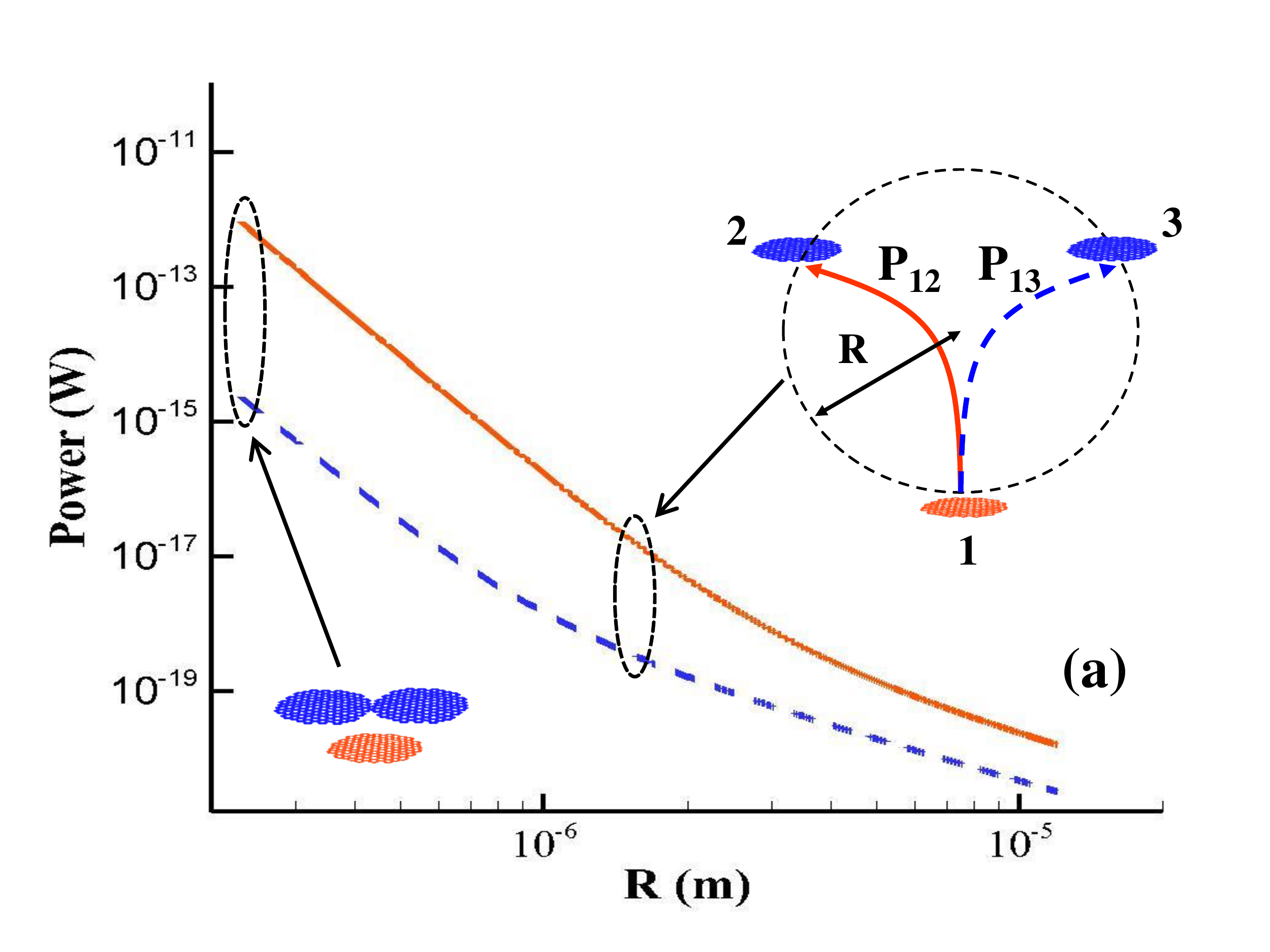}
  \includegraphics[scale=0.35]{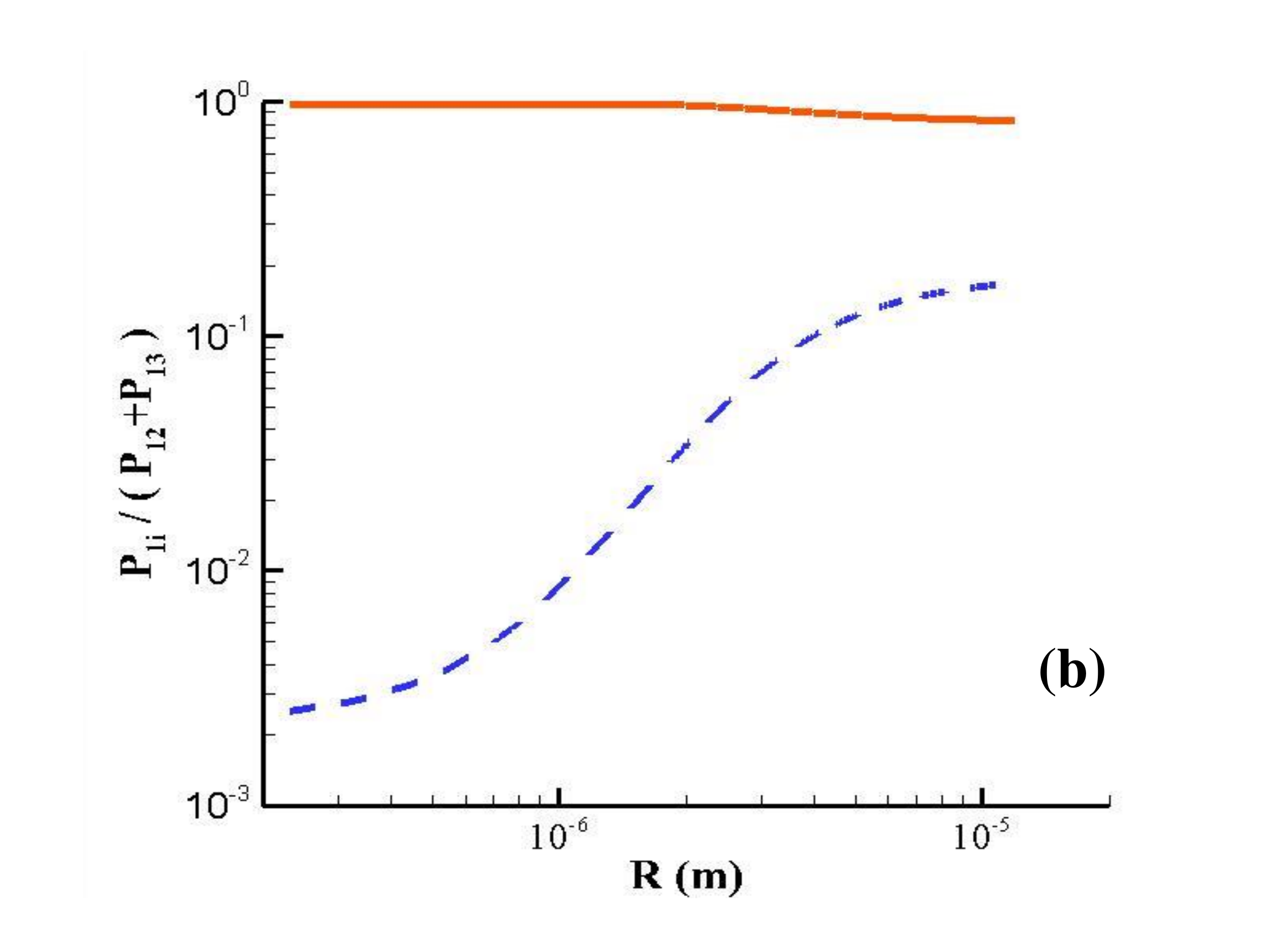}
   \caption{(a) Thermal power exchanged in the near-field between graphene disks 100nm 
            radius versus the separation distance in a three body system. 
            The Fermi level of disks 1, 2 and 3 are $E_{F_1}=0.1\,{\rm eV}$,  $E_{F_2}=0.1\,{\rm eV}$ 
            and  $E_{F_3} = 0.8\,{\rm eV}$ while their temperatures are $T_{1}=350\,{\rm K}$ and $T_{2} = T_{3} = 273\,{\rm K}$. 
            All disks are parallel to the $x^{(0)}$-$y^{(0)}$ plane and $\epsilon_h=1$.
            (b) Plot of $\mathcal{P}_{12}/(\mathcal{P}_{12} + \mathcal{P}_{13})$ and $\mathcal{P}_{13}/(\mathcal{P}_{12} + \mathcal{P}_{13})$ for 
            the same configuration.}
\end{figure}

The radiative power exchanged between the nano-disks can be described using the many-body radiative heat transfer 
theory~\cite{PBAEtAl2011,Riccardo} extended to anisotropic structures~\cite{Nikbakht,Incardone}. By neglecting 
the far-field radiative losses towards the surrounding, which is typically small compared to the near-field contribution, 
the net power exchanged in the near-field between the $i^{th}$ and the $j^{th}$ nano-disk reads~\cite{PBAEtAl2011} 
\begin{equation}
  \mathcal{P}_{ij}=3\int_{0}^{\infty}\frac{\rd\omega}{2\pi}\,[\Theta(\omega,T_{i})-\Theta(\omega,T_{j})]\mathcal{T}_{i,j}(\omega)\label{Eq:InterpartHeatFlux},
\end{equation}
where $\Theta(\omega,T)={\hbar\omega}/[{e^{\frac{\hbar\omega}{k_B T}}-1}]$ is the mean energy of a harmonic oscillator in
thermal equilibrium. $\mathcal{T}_{i,j}$ denotes the transmission coefficient between the  $i^{th}$ and the $j^{th}$ nano-disk 
and is defined as
\begin{equation}
  \mathcal{T}_{i,j}(\omega)=2\Im\Tr\bigl[\mathds{A}_{ij}\Im\boldsymbol{\chi}_j\mathds{C}_{ij}^{\dagger}\bigr],
\end{equation}
where $\boldsymbol{\chi}_j$, $\mathds{A}_{ij}$ and $\mathds{C}_{ij}$ are the susceptibility tensor plus two 
matrices which are given in terms of Green's tensors and the polarizability tensors. These quantities are 
defined in Ref.~\cite{Nikbakht}.

Let us now apply this theoretical framework to describe the operating modes of near-field heat flux splitter. 
To this end, we consider a triplet of equal graphene disks of diameter $D$ as depicted in the inset of Fig.~2. 
The center of mass of the first disk is located in the origin of the canonical frame while the centers of mass of 
the two other disks are located on a virtual sphere of radius $R$ which is resting on the $x^{(0)}$-$y^{(0)}$ plane
such that its contact point coincides with the origin. For convenience we consider only the situation, where all
centers of masses are bound to the $x^{(0)}$-$z^{(0)}$ plane and equi-distant to each other. Hence, the distance
between the mass centers of the disks is  $l=\frac{\sqrt3}{2} R$.

Now, let us first consider the case, where all disks are parallel to the  $x^{(0)}$-$y^{(0)}$ plane.
Furthermore, we assume that $T_{1}=350\,{\rm K}$ and $T_{2} = T_{3} = 273\,{\rm K}$ so that there is a heat
flux from disk $1$ (see Fig.~2) to disk $2$ and $3$. If the Fermi levels of all three disks are the same then 
the power $\mathcal{P}_{12}$ and $\mathcal{P}_{13}$ exchanged between disk $1$
and $2$ and between $1$ and $3$, respectively, will be the same as well. This situation corresponds to the trivial 
operating mode of the heat flux splitter as a 50:50 splitter. 

On the other hand, if we change the Fermi level of disk $3$ to $E_{F,3}=0.8\,{\rm eV}$, we have an asymmetric situation. 
The resulting powers $\mathcal{P}_{12}$ and $\mathcal{P}_{13}$ are plotted in 
Fig.~2 versus the radius $R$ of the virtual sphere choosing a disk radius of $D/2=100\,{\rm nm}$. 
First of all, we see that $\mathcal{P}_{12}$ can be almost three orders of magnitude larger 
than $\mathcal{P}_{13}$ which means that nearly the full heat flux is directed from 
disk $1$ towards disk $2$. Here the heat flux splitter acts as a 99:1 beam splitter for distances around on micron. 
Clearly, by a gradual change of the Fermi level all values between 50:50 and 99:1 are attainable.
Regarding the decay of the fluxes with $R$ we find that in the far-field $\mathcal{P} \propto 1/R^2$  and in the near-field regime
$\mathcal{P} \propto 1/R^2$ expected from the dipole-dipole interaction, where the field radiated by each 
disk decays as $1/R$ or $1/R^3$ in far- or near-field regime, respectively. Of course, the direction of the
heat flow can be directed towards disk 3 by interchanging the Fermi levels of disk $2$ and $3$ dynamically by electrical 
gating, for instance.

\begin{figure}[Hhbt]
   \includegraphics[scale=0.4]{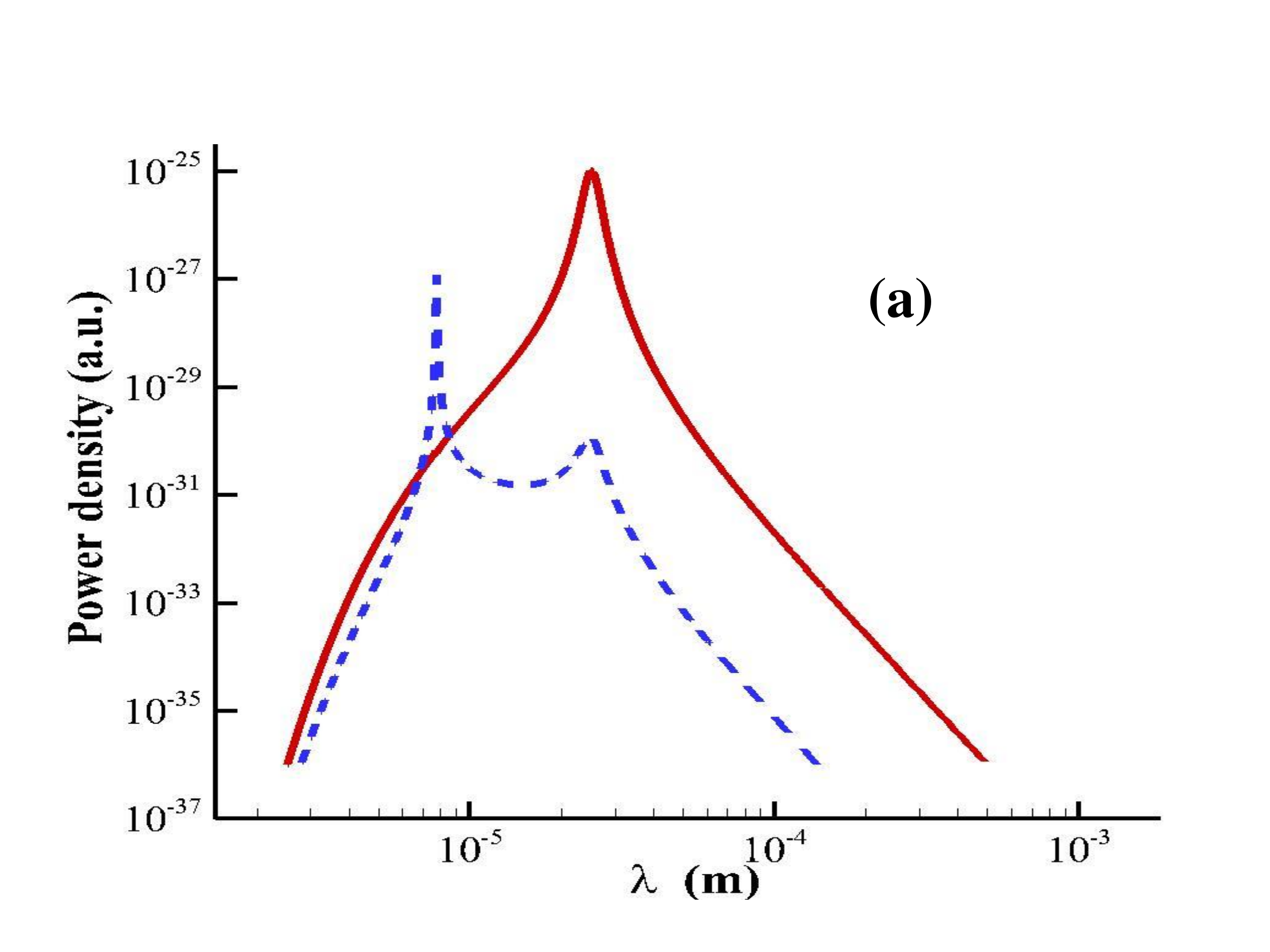}
\includegraphics[scale=0.35]{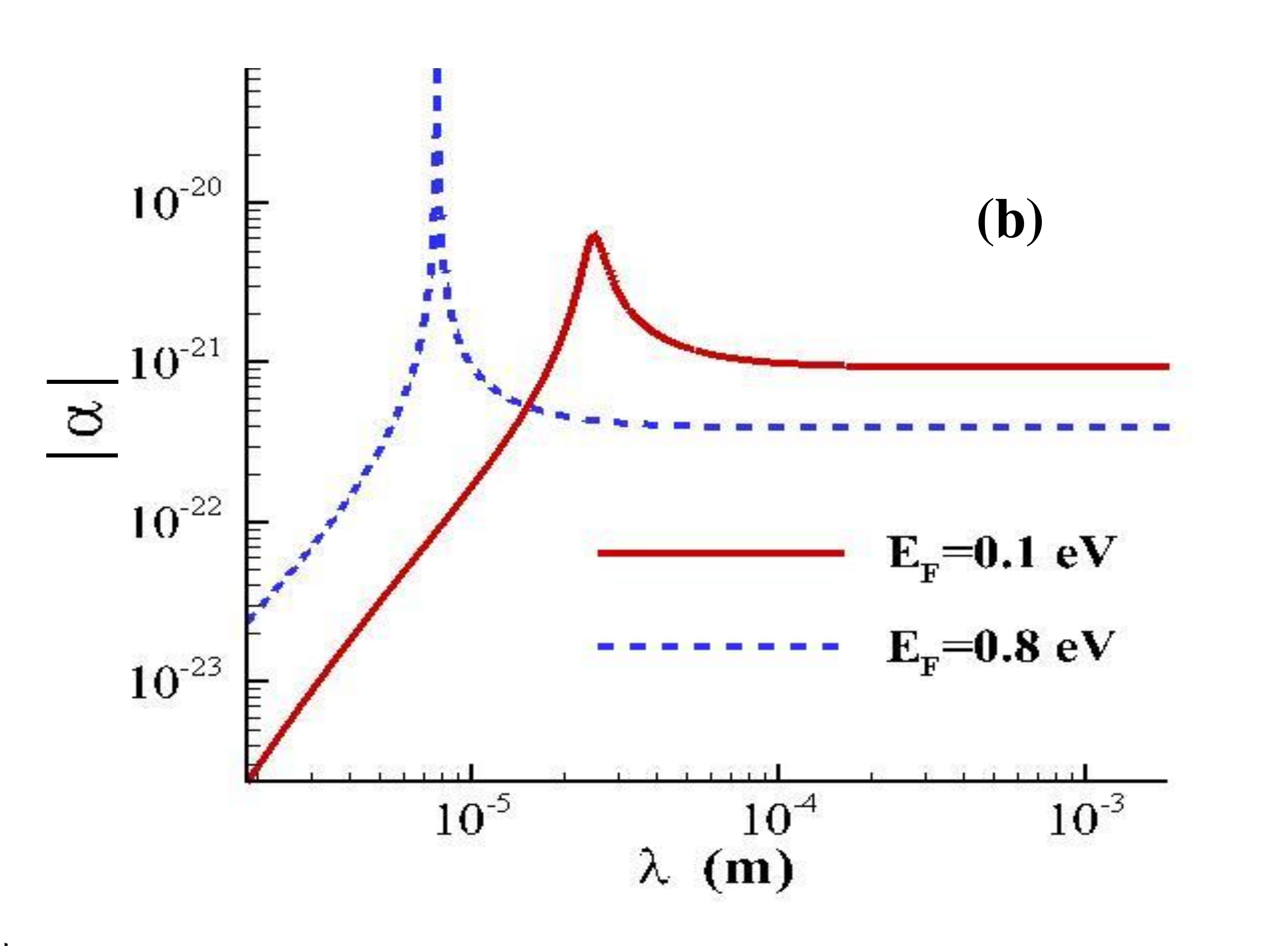}
   \caption{(a)  Spectral power exchanged between the graphene disks $1$ and $2$ (red) and between $1$ and $3$ (dashed blue) in near-field for 
            the case considered in Fig.~2. (b) Modulus of the polarizability of the nano-disks from Eq.~(\ref{Eq:Polarizability}) 
            for the two different Fermi levels $E_F$.}
\end{figure}

To explain the huge difference of magnitude of the power echanged between two graphene nano-disks
with identical Fermi levels and that between two disks having different Fermi levels, we analyze the 
spectra of the power fluxes $\mathcal{P}_{12}$ and $\mathcal{P}_{13}$
calculated in Fig.~2. These spectra are plotted in Fig.~3(a) for the radius $R=2D/\sqrt3$ with $D=200\,{\rm nm}$, i.e.\ 
$R = 231\,{\rm nm}$. At small wavelengths, we see that both spectra are very similar. In contrast, at larger 
wavelengths, the shape of spectra are radically different. Excepted in a narrow spectral region centered around
the thermal wavelengths $\lambda_{T_i}=c\hbar/(k_B T_i)$ (i.e.\ spectral region where the Planck function is maximal), 
the spectrum of the exchanged power between two graphene nano-disks with different Fermi levels is more than two 
orders of magnitude smaller than that for two nano-disks with identical Fermi levels. Furthermore, the spectra show
sharp resonances. These resonances are located exactly at wavelengths where the polarizability of the graphene disks 
become resonant (see Fig.~3(b)). These resonances are due to the localized plasmon polaritons in graphene. 
At  $E_{F}=0.1\,{\rm eV}$ this localized plasmon is located in the midinfrared at $\lambda=25.1\, \mu{\rm m}$. When the 
density of electron increases in graphene this resonance shifts to smaller wavelengths so 
that at $E_{F}=0.8\,{\rm eV}$ it is located in the infared at $\lambda=7.75\mu m$. When using identical Fermi levels
these resonances overlap perfectly resulting in a large heat flux, while when using different Fermi levels
this overlap becomes smaller resulting in smaller heat fluxes. Hence, we see that, by changing the Fermi level of 
graphene, we can switch the different channels for the heat transport in this nanostructure network on and off.  

Finally, we show on Fig.~4 the angular dependence of heat flux exchanged between the disks for different separation 
distances. We observe that the exchanged power $\mathcal{P}_{12}$ decays monotically when $\theta$ grows, whereas
$\mathcal{P}_{13}$ increases monotonically. Moreover, when the mismatch of the Fermi levels in the graphene nano-disks 
is large the angular dependence is weak, whereas when the  Fermi levels are identical the 
angular dependence becomes more important.  

\begin{figure}[Hhbt]
 \includegraphics[scale=0.35]{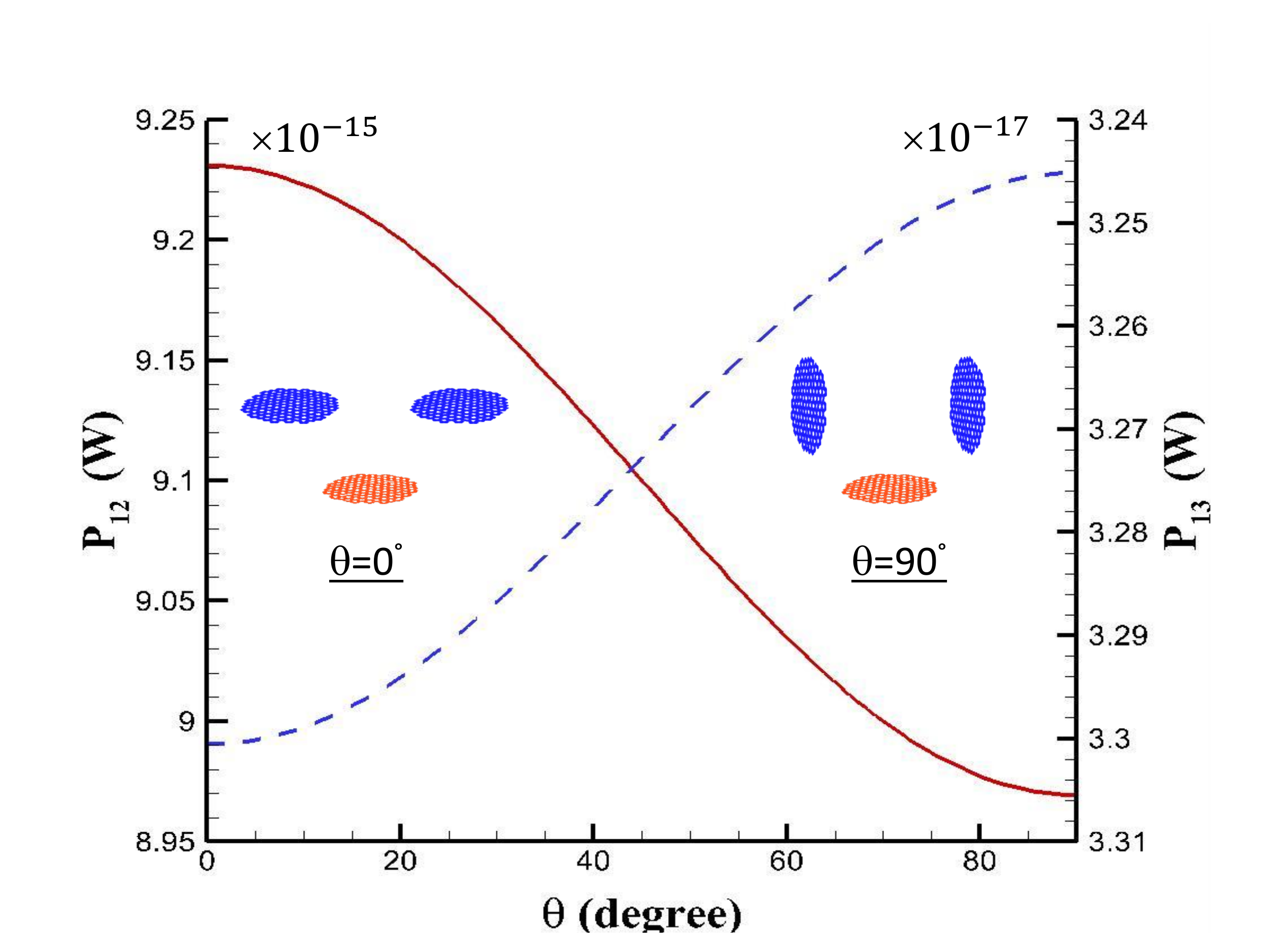}
 \caption{Angular depedence of heat power $\mathcal{P}_{12}$ (red) and  $\mathcal{P}_{13}$ (dashed blue) 
          exchanged in the near-field between graphene disks of $100\,{\rm nm}$ radius in the heat flux splitter. 
          The Fermi level of disks 1, 2 and 3 are $E_{F_1}=0.1\,{\rm eV}$,  $E_{F_2}=0.1\,{\rm eV}$ and  $E_{F_3}=0.8\,{\rm eV}$ 
          while their temperatures are $T_{1}=350\,{\rm K}$, $T_{2}=T_{3}=273\,{\rm K}$. The radius of the virtual sphere is $R=500\,{\rm nm}$.}
\end{figure}

In conclusion, we have introduced the basic building blocks for a control of heat flow directions at the nanoscale 
by introducing a heat flux splitter based on a graphene nano-disk network. The working principle relies on the
near-field interaction of the localized plasmons in graphene nano-disks which can be tuned dynamically by electrical 
means. In this Letter, we have limited ourself to the description of the simplest configuration for a heat flux 
splitter. However, the basic concepts introduced here could be easily generalized to deal with heat transfers 
in much more complex plasmonic networks. In such networks the near-field heat flux splitting could be used to 
dynamically manage heat exchanges in integrated architectures of nano-objects.

 P.B.-A. acknowledges discussions with M. Nikbakht and J. P. Hugonin.

\end{document}